# Superconductivity in Scandium Borocarbide with orbital hybridization

W. Wu†[1,2], Y. J. Li†[1,2], J. H. Zhang[1,2], Z. H. Yu[3], Z. Y. Liu[1,2], P. Zheng[1,2], H. X. Yang[1,2], C. Dong[1,2,6], K. Liu*[4], T. Xiang*[1,2,5], and J. L. Luo*[1,2,6]

[1]*Beijing National Laboratory for Condensed Matter Physics and Institute of Physics, Chinese Academy of Sciences, Beijing 100190, China*

[2]*School of Physical Sciences, University of Chinese Academy of Sciences, Beijing 100190, China*

[3]*School of Physical Science and Technology, ShanghaiTech University, Shanghai 201210, China*

[4] *Department of Physics and Beijing Key Laboratory of Opto-electronic Functional Materials & Micro-nano Devices, Renmin University of China, Beijing 100872, China*

[5]*Kavli Institute for Theoretical Sciences, Beijing 100190, China*

[6]*Songshan Lake Materials Laboratory, Dongguan, Guangdong 523808, China*

†W. Wu, Y. J. Li, contributed equally to this work
\*Correspondence should be addressed to K. Liu (*kliu@ruc.edu.cn*) and
T. Xiang (*txiang@iphy.ac.cn*), J. L. Luo (*jlluo@iphy.ac.cn* )

## ABSTRACT

Exploration of superconductivity in light element compounds has drawn considerable attention because those materials can easily realize the high $T_C$ superconductivity, such as LnNi$_2$B$_2$C ($T_C$ =17 K), MgB$_2$ ($T_C$ =39 K), and very recently super-hydrides under pressure ($T_C$ =250 K). Here we report the discovery of bulk superconductivity at 7.8 K in scandium borocarbide Sc$_{20}$BC$_{27}$ with a tetragonal lattice which structure changes based on the compound of Sc$_3$C$_4$ with very little B doping. Magnetization and specific heat measurements show bulk superconductivity. An upper critical field of $H_{c2}(0) \sim 8$ T is determined. Low temperature specific-heat shows that this system is a BCS fully gapped s-wave superconductor. Electronic structure calculations demonstrate that compared with Sc$_3$C$_4$ there are more orbital overlap and hybridization between Sc 3$d$ electrons and 2$p$ electrons of C-C(B)-C fragment in Sc$_{20}$BC$_{27}$, which form a new electric conduction path of Sc-C(B)-Sc. Those changes influence the band structure at the Fermi level and may be the reason of superconductivity in $Sc_{20}BC_{27}$.

## I. INTRODUCTION

High $T_C$ superconductor may be anticipated for hydrides, carbides, and borides due to their light masses and strong covalent bonding, which yields high vibrational frequencies. From the McMillan relation $kT_C = 1.13\hbar\omega\exp[-1/N(E_F)V]$ [1], where $V$ is a measure of the electron-phonon coupling and $\omega$ is a characteristic phonon frequency with the similar magnitude to the Debye frequency. On the basis of this expression, large values of either $N(E_F)$ (A15 compounds) or $V$ (light elements : borides, carbides) or both lead to high $T_c$ values, such as for $LnNi_2B_2C$ ($T_c$ = 17 K) [2-4], $MgB_2$ ($T_C$ = 39 K) [5], $Y_2C_3$ ($T_c$ = 18 K) [6], and the recently synthesized hydrides at mega-bar pressures, in which nearly room-temperature superconductivity has been realized in $H_3S$ ($T_c$ = 203 K) and $LaH_{10-x}$ ($T_c$ > 260 K) [7-8]. Metal carbides are good candidates to explore high $T_c$ superconductors and they also provide a bridge which links the organic and inorganic materials [9]. There are many carbide compounds containing C2 fragments that are superconductors, such as $Y_2C_3$ (18 K) [6], $YC_2$ (3.9 K) [10], $Y_2C_2I_2$ (10 K) [11], whose structures contain molecular anionic fragments, (C-C) or (B-C-B), that interact with a metal framework.

$Sc_3C_4$ is the first model compound containing $C_3^{4-}$ fragments and some other examples are $Mg_2C_3$ [12] and $Ca_3Cl_2C_3$ [13]. $Sc_3C_4$ crystallizes in a tetragonal structure (Z = 10, P4/mnc, $a$ = 748.73(5) pm, $c$ = 1502.6(2) pm) as shown in Fig. 1(a). The unit cell contains eight C3 fragments, two C2 fragments and twelve isolated C atoms, with the $S_{C30}(C_3)_8(C_2)_2(C)_{12}$ composition [14-17]. $Sc_3C_4$ is a metallic conductor and a Pauli paramagnet, but is not a superconductor. Oyama analyzed the electronic band structures of $Sc_3C_4$ and found that it is Sc $3d$ orbitals instead of the $p$ orbitals of C2 and C3 that provide large contribution at $E_F$ [18]. There is no sufficient overlap between C $p$ orbitals and Sc $3d$ orbitals around $E_F$ and the calculated spectra of the molecular fragments do not meet the energy requirements as found in $LnC_2$, $Ln_2C_3$, and $Ln_2C_2X_2$ superconductors [18]. It is worth trying to change the electronic structure of $Sc_3C_4$ via doing elements or applying pressure to induce superconductivity.

Here we report the crystal structure and basic superconducting properties of the new superconductor $Sc_{20}BC_{27}$, whose structure has an adjustment on the base of $Sc_3C_4$ by doping very little B element. Our powder X-ray diffraction (PXRD) characterization agreed with Rietveld fitting that $Sc_{20}BC_{27}$ crystallizes in the space group $P4/ncc$ (No. 130). The angle of C-C(B)-C decrease from 175.8° in $Sc_3C_4$ to 165.2° in $Sc_{20}BC_{27}$. Temperature-dependent electrical resistivity, magnetic susceptibility, and specific heat measurements were used to characterize the superconductivity at $T_c$ = 7.8 K. Low temperature specific-heat shows this system is a BCS fully gapped s-wave superconductor. Electronic structure calculations demonstrate that there are more orbital overlap and hybridization between $p$ electrons of middle C(B) atom from C-C(B)-C fragment and Sc $3d$ electrons in $Sc_{20}BC_{27}$ compared with $Sc_3C_4$ around the Fermi level, which may be the reason of superconductivity in $Sc_{20}BC_{27}$.

We synthesized and confirmed the superconductivity on the base of doping little B in $Sc_3C_4$ in 2018 [19]. Recently, Ninomiya and coworkers have independently obtained similar results on the occurrence of superconductivity through resistivity and susceptibility measurements and confirmed the final crystal structure of this superconductor [20]. Now we found our result of structure is the same with that of Ninomiya *et al.*, indicating that they are the same compound.

## II. EXPERIMENTAL AND THEORETICAL METHODS

The starting materials for the synthesis of polycrystalline $Sc_3C_4$ were graphite (C 99.99%), amorphous boron (B 99.95) and piece of scandium (Sc 99.99%). The Sc and C, B chunks were weighed out in a 3:4:0.1 ratio. Then C and B powder was pressed into a pellet and put together with piece of Sc, arc-melted to have a metal chunk for subsequent meltings, and the samples were arc-melted four times in Ar atmosphere of 500 mbar. In between each melting the arc-melted buttons were flipped to ensure homogeneous samples.

The resulting crystals were characterized by X-ray diffraction (XRD) with Cu K$_{\alpha 1}$

radiation at room temperature. The XRD pattern of $Sc_{20}BC_{27}$ was analyzed with the GSAS program with a user interface EXPGUI [21]. We carried out the specific heat, DC susceptibility and electrical resistivity measurements in a Quantum Design Physical Property Measurement System (PPMS-9), respectively.

We conducted first-principles electronic structure calculations on $Sc_{20}BC_{27}$ and $Sc_3C_4$ using the projector augmented wave (PAW) method [22] as implemented in the VASP package [23-24]. The generalized gradient approximation (GGA) of Perdew-Burke-Ernzerhof (PBE) type was adopted for the exchange-correlation functional [25]. The plane-wave basis set with a kinetic energy cutoff of 520 eV was employed. For the Brillouin zone sampling of the tetragonal cells of $Sc_{20}BC_{27}$ and $Sc_3C_4$ ($Sc_{30}C_{40}$), the $6 \times 6 \times 4$ and $6 \times 6 \times 3$ $k$-point meshes were used, respectively. The doping of B atoms in the C-C-C fragment was investigated with the virtual crystal approximation (VCA) approach. The density of states (DOS) was calculated by using the tetrahedron method with Blöchl corrections.

### III. RESULTS AND ANALYSIS

Figure 1(c) shows the result of GSAS refinement of $Sc_{20}BC_{27}$ under ambient condition, which indicates that $Sc_{20}BC_{27}$ crystallized in a tetragonal structure with space group $P4/ncc$ (#130). The lattice parameters of $Sc_{20}BC_{27}$ are $a$ = 7.5188(1) Å and $c$ = 10.0473(4) Å. Comparison with structure of $Sc_3C_4$ in Fig. 1(a), there are some minor changes in the structure of $Sc_{20}BC_{27}$ [Fig. 1(b)]. Most of isolated carbon atoms and NaCl-type lattice containing C2 have disappeared, but C3 fragment and NaCl-type lattice containing C1 keep at the tetragonal structure. The lattice parameter $a$ has a little change, but the lattice parameter $c$ reduces by 30%. Meanwhile, B atom occupies the middle position of C3 fragment, and the angle of C-C-C in $Sc_3C_4$ decreases from 175.8° to 165.2° for C-C(B)-C in $Sc_{20}BC_{27}$.

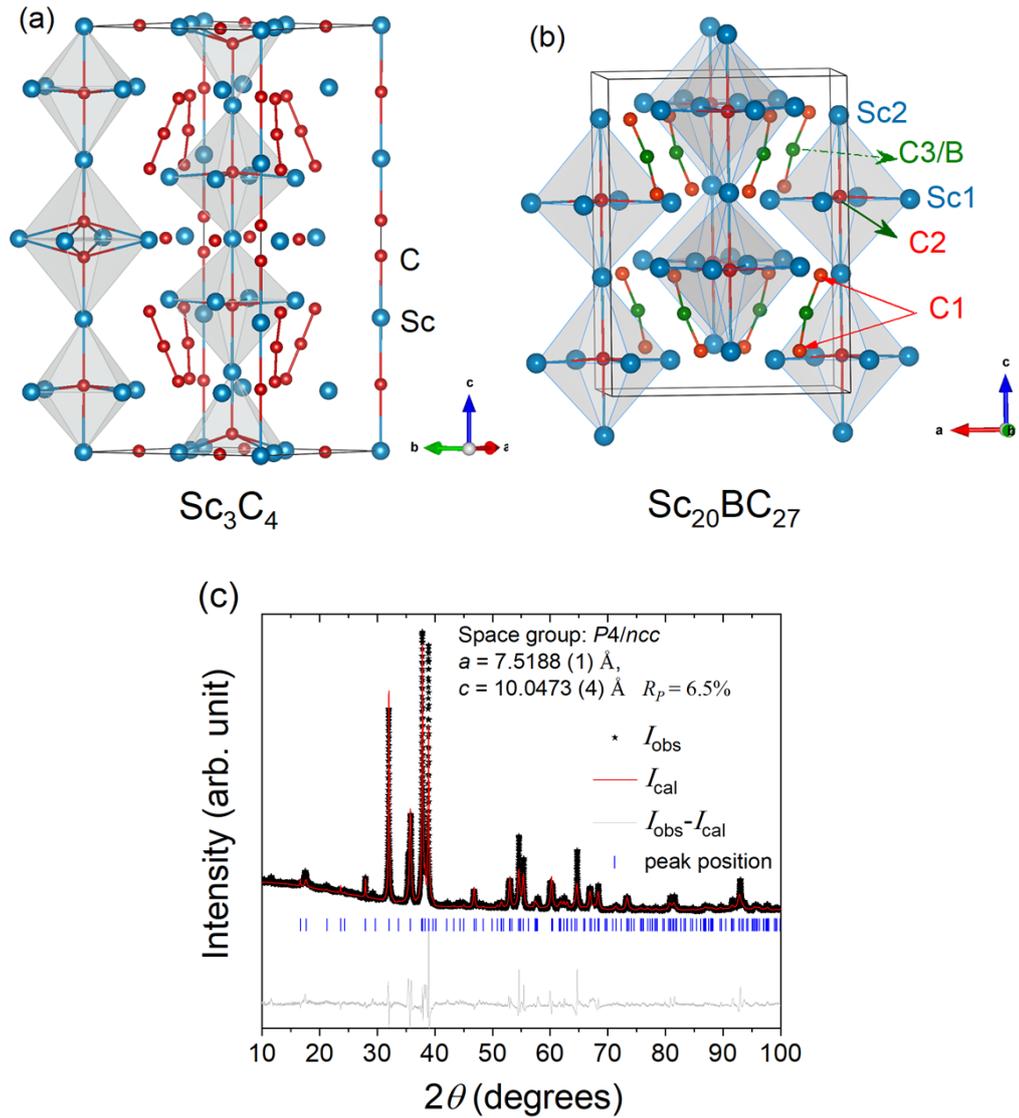

**Fig. 1.** Crystal structures of (a) $Sc_3C_4$ and (b) $Sc_{20}BC_{27}$. (c) Typical Rietveld refinement of $Sc_{20}BC_{27}$ under ambient conditions. The vertical bars represent the calculated Bragg reflection positions of the diffraction peaks for $Sc_{20}BC_{27}$. The difference between the observed (scatters) and the fitted patterns (line) is shown at the bottom of the diffraction peaks.

Figure 2 shows the temperature dependence of electrical resistivity for $Sc_3C_4$ and $Sc_{20}BC_{27}$ from 2 K to 300 K. $Sc_3C_4$ is a metallic conductor, but is not a superconductor. The slight upturn of the $\rho$-T curve was observed. The magnetoresistance (MR) of $Sc_3C_4$ is 14% at 2 K under 9 T (as shown in the inset of Fig. 2). For the $Sc_{20}BC_{27}$, the normal state resistivity is metallic, and there also exists an upturn of the $\rho$-T curve below ~30 K. At the low temperatures, a sharp superconducting transition shows up at about 7.8

K and the zero resistivity appears below 7.3 K. The width of superconducting transition is 0.5 K, indicating the poor crystallinity in $Sc_{20}BC_{27}$.

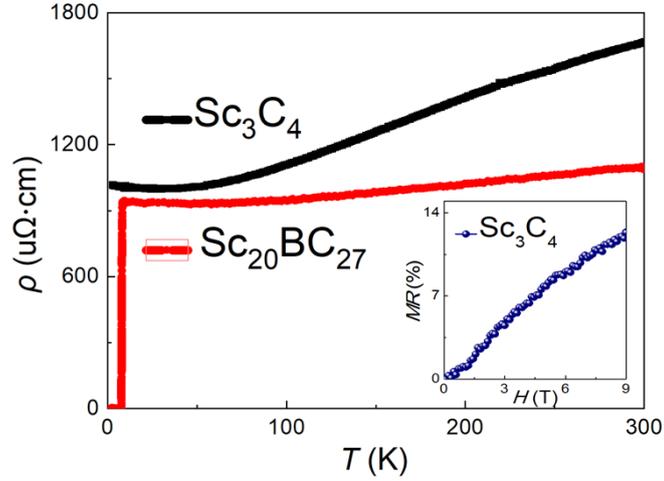

**Fig. 2.** (a) Temperature dependence of the resistivity for $Sc_3C_4$ and $Sc_{20}BC_{27}$ at zero field. The inset shows magnetoresistance (MR) of $Sc_3C_4$ at 2 K.

Figure 3 shows $\rho$(T) curves of $Sc_{20}BC_{27}$ under various magnetic fields up to 9T. The width of superconducting transition in resistivity changes slightly with increasing magnetic field. The zero-temperature upper critical field $H_{c2}(0)$ is about 8 T, which is relatively large compared with the type-I BCS superconductors. The upper critical field $H_{c2}(0)$ can be estimated with the formula $H_{c2}(T) = H_{c2}(0)(1 - t^2)$, where t is the reduced temperature t = $T/T_c$. By fitting, we obtained $H_{c2}(0) \sim 8.2$ T with the Werthamer-Helfand-Honenberg (WHH) formula $\mu_0 H_{c2}(0) = -0.693 T_c \left( \frac{d\mu_0 H_{c2}(T)}{dT} \bigg|_{T_c} \right)$ [26]. We estimated the Ginzburg-Landau coherence length $\xi_{GL}(0)$ from the upper critical field $\mu_0 H_{c2}(0)$ according to the relation $\mu_0 H_{c2}(0) = \Phi_0 / 2\pi \xi_{GL}^2$ [27], where $\Phi_0$ is the quantum flux h/2e, and we got $\xi_{GL}(0) = 61$ Å.

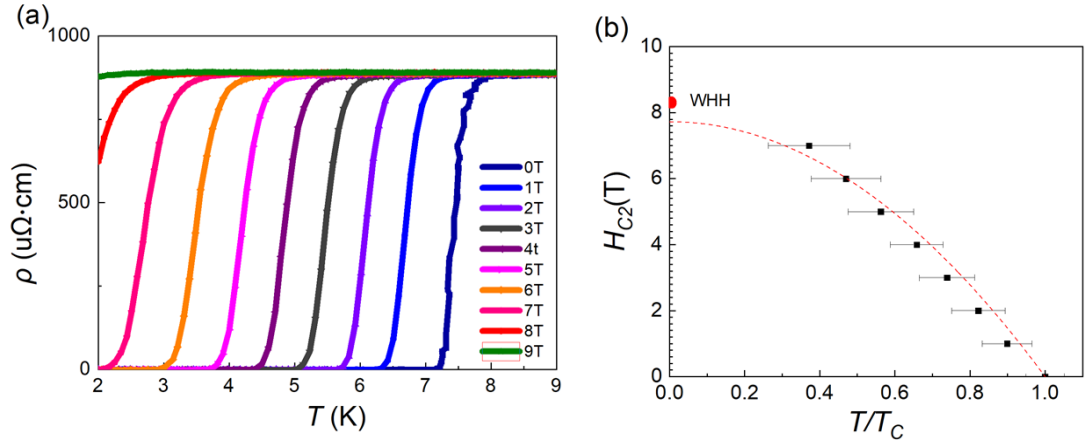

**Fig. 3.** (a) Temperature dependence of the resistivity at different fields. (b) Temperature dependence of the upper critical magnetic field. The red solid circle shows the upper critical field obtained from WHH fitting.

To further confirm the bulk superconductivity, we resorted to DC magnetic susceptibility measurements in $Sc_{20}BC_{27}$. Figure 4 shows zero field cooling (ZFC) and field cooling (FC) processes of the susceptibility $\chi$ in the low temperature under the applied field of H = 10 Oe. Diamagnetic signal is probed below $T_c$ due to superconducting transition and tends to saturate at low temperatures. The shielding fraction estimated from ZFC data at 2 K is about 80%.

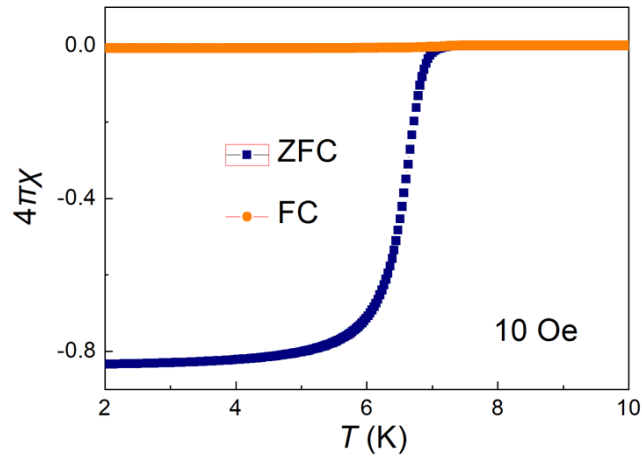

**Fig. 4** Temperature dependence of the magnetic susceptibility (M/H) measured in the applied field of H = 10 Oe. ZFC and FC indicate the zero-field-cooling and field-cooling processes, respectively.

Figure 5(a) shows the specific heat coefficient $C/T$ versus temperature square $T^2$ at zero field. The character of superconductivity is also verified by an obvious peak at $T_c$ = 7.8 K, matching well with the resistivity and susceptibility measurements. Above $T_c$, $C/T$ shows linear $T^2$ dependence indicating the specific heat $C$ could be well fitted by $C$

= $\gamma T + \beta T^3$, where the first term $\gamma T$ is the contribution from the conduction electrons ($\gamma$ is Sommerfeld coefficient) and the second term $\beta T^3$ is the contribution from the phonon part. The best fit gives $\gamma$ = 53 mJ/mol.K$^2$ and $\beta$ = 0.67 mJ/mol.K$^4$. The corresponding Debye temperature is obtained from $\beta$ to be $\Theta_D$ = 550 K. At low temperatures $C/T$ extrapolates to a value of $\gamma_0$ = 15mJ$^{-1}$mol K$^2$, indicating there is residual density of states at the Fermi energy at zero temperature. From the difference electronic specific heat coefficient ($\Delta C_e/T$) between the superconducting and normal states is plotted as in the inset of Fig. 5(a), the normalized specific heat jump at $T_c$ is found to be $\Delta C/(\gamma-\gamma_0)T_c$ = 1.35, which is close to the BCS prediction for weak-coupling superconductivity of 1.43 [27].

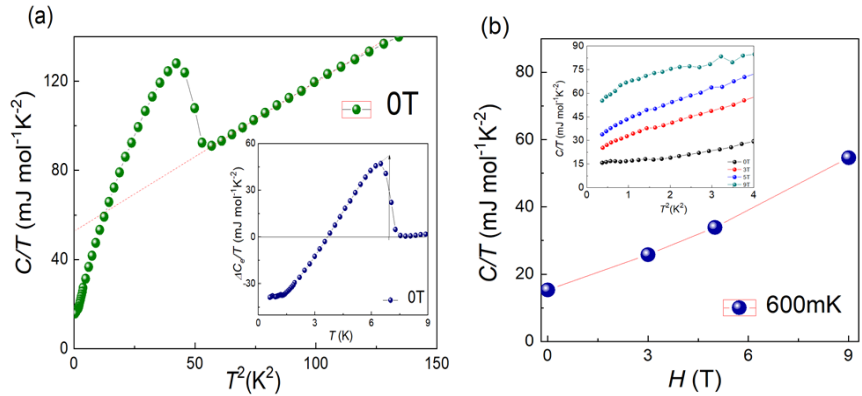

**Fig. 5.** (a) Specific heat coefficient $C/T$ of Sc$_{20}$BC$_{27}$ as a function of $T^2$. The inset shows temperature dependence of the difference electronic specific heat coefficient ($\Delta C_e/T$) between the superconducting and normal states. (b) Magnetic field dependence of the field induced specific heat coefficient $C/T$ ($H$) at 600mK. The inset shows the specific heat coefficient $C/T$ of Sc$_{20}$BC$_{27}$ at different field in the low temperature

Figure 5(b) shows the field dependence of the electronic specific heat $C/T$ at 600 mK. The value is obtained from the inset of Fig. 5(b) which shows the temperature square dependence of the electronic specific heat $C/T$ at different magnetic field in the low temperature. It is clear that $C/T$ is close to a linear relation which indicates again the presence of isotropous gap.

In order to understand the superconductivity observed in Sc$_{20}$BC$_{27}$, it is necessary to compare the changes of crystal structures and electronic structures from Sc$_3$C$_4$ to Sc$_{20}$BC$_{27}$. First, the (C-C-C) fragment has a big distortion by doping B element that the angle decreases about 10° from 175.8° to 165.2° [Fig. 6(e)]. This distortion also occurs in the carbide LaNi$_2$B$_2$C where B-C-B $\pi$-nonbonding orbital is tuned by the position of the La d$x^2$-$y^2$ orbital, which leads to second order Jahn-Teller instabilities [19]. But in other systems with (C-C(B)-C) fragment, the angle has little change, for instance, it is 174.6° in La$_5$B$_2$C$_{60}$ [28] and 174.3° in Y$_{15}$B$_4$C$_{14}$ [29].

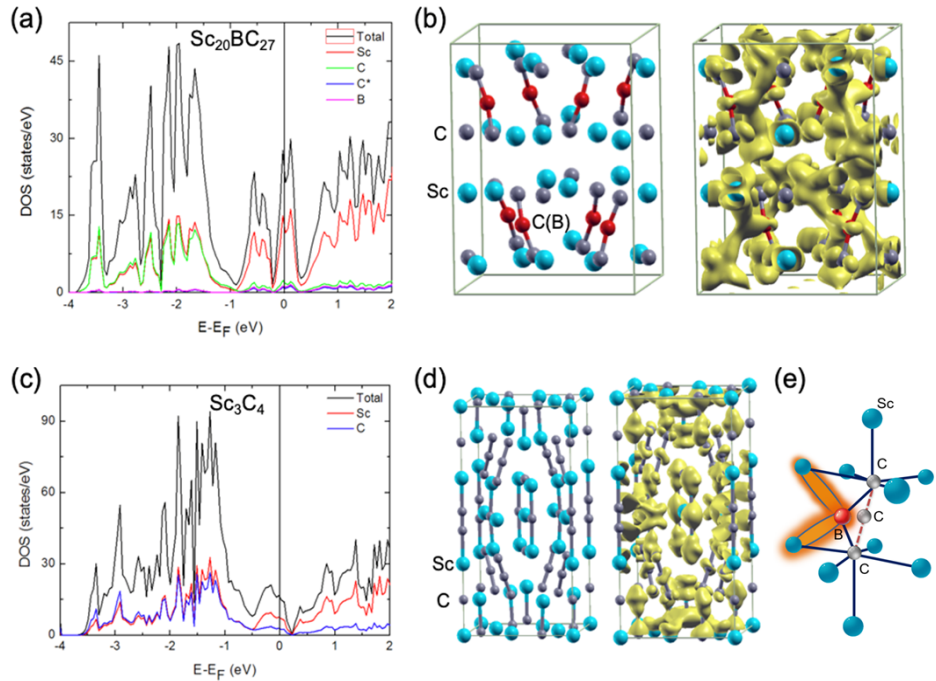

**Fig. 6.** Calculated electronic density of states for (a) Sc$_{20}$BC$_{27}$ and (c) Sc$_3$C$_4$. 'C*' and 'B' denote the middle C and B atoms in the C-C(B)-C fragment of Sc$_{20}$BC$_{27}$. Crystal structure and integrated charge densities in an energy interval [-0.25, 0] eV for (b) Sc$_{20}$BC$_{27}$ and (d) Sc$_3$C$_4$. Here the Fermi energy E$_F$ sets to zero. (e) Schematic coordination of C-C(B)-C fragment.

Second, we performed first-principles electronic structure calculations on Sc$_{20}$BC$_{27}$ and Sc$_3$C$_4$. The calculated electronic density of states (DOS) for Sc$_{20}$BC$_{27}$ in Fig. 6(a) demonstrate that in comparison with C and B atoms, the Sc orbitals have major contributions around the Fermi level (E$_F$). Moreover, there is a sharp peak at E$_F$ and a deep dip at about -0.25 eV below E$_F$. By integrating the charge densities in the energy interval of [-0.25, 0] eV referring to E$_F$, we learn that the 3$d$ orbitals of Sc atoms and the 2$p$ orbitals of middle C(B) atoms in the C-C(B)-C fragment of Sc$_{20}$BC$_{27}$ have strong hybridization, forming an electric conduction path of Sc-C(B)-Sc [as shown in Fig. 6(b)]. Instead, the DOS of Sc$_3$C$_4$ around E$_F$ are broader compared with that of Sc$_{20}$BC$_{27}$

[Fig. 6(c)], and there is no obvious orbital hybridization between Sc atoms and the middle C atoms in the C-C-C fragment [Fig. 6(d)]. The distinction in the electronic structures of $Sc_{20}BC_{27}$ and $Sc_3C_4$ can be well understood by the schematic coordination of C-C(B)-C fragment in Fig. 6(e), where the distorted C-C(B)-C fragment in $Sc_{20}BC_{27}$ with B doping facilitates the overlap of C(B) $2p$ orbitals and Sc $3d$ orbitals. The occurrence of superconductivity in $Sc_{20}BC_{27}$ may thus be due to the distortion of (C-C-C) fragment and the resulted change in electronic structures near the Fermi level.

## IV. CONCLUSION

We report the experimental results for a polycrystalline sample of ternary borocarbide $Sc_{20}BC_{27}$. $Sc_{20}BC_{27}$ were successfully synthesized using arcing method. The Rietveld refinements demonstrate that $Sc_{20}BC_{27}$ has a tetragonal lattice structure without C-C fragment compared with $Sc_3C_4$. Bulk superconductivity with $T_c \sim 7.8$ K is observed from the resistivity, susceptibility, and specific heat measurements. Low temperature specific heat data shows the superconductivity is $s$-wave pairing symmetry. The specific heat shows linear relation with magnetic field, suggesting the existence of isotropous gap. Electronic structure calculations demonstrate that there is more hybridization between the $p$ orbitals of middle B(C) atoms in the C-B(C)-C fragment and the $3d$ orbitals of Sc atoms in $Sc_{20}BC_{27}$ compared with those in $Sc_3C_4$, which forms a new electric conduction path of Sc-B(C)-Sc. Those changes influence the electronic structure at the Fermi level and may be the reason of superconductivity in $Sc_{20}BC_{27}$. Further experimental studies are needed to increase the superconducting transition temperature via doping other elements or applying high pressure.

## ACKNOWLEDGEMENT

We wish to thank Miao Gao and Pei-Jie Sun for helpful discussions. This work was supported by the National Basic Research Program of China (Grants No. 2014CB921500 and No. 2017YFA0302900), the National Science Foundation of China (Grants No. 11674375, No. 11634015, and No. 11774424), the Strategic Priority Research Program and Key Research Program of Frontier Sciences of the Chinese Academy of Sciences (Grant Nos. XDB07020200), the CAS Interdisciplinary

Innovation Team, the China Postdoctoral Science Foundation, the Fundamental Research Funds for the Central Universities, and the Research Funds of Renmin University of China (Grant No. 19XNLG13). Computational resources were provided by the Physical Laboratory of High Performance Computing at Renmin University of China.